\documentclass[12pt,a4paper,final]{iopart}
\pdfoutput=1
\usepackage{iopams}  
\usepackage{graphicx}
\usepackage[breaklinks=true,colorlinks=true,linkcolor=blue,urlcolor=blue,citecolor=blue]{hyperref}
\usepackage{xcolor}

\begin{document}
\title{Driven Tracers in a One-Dimensional Periodic Hard-Core Lattice Gas}
\author{Ivan Lobaskin and Martin R. Evans}
\address{School of Physics and Astronomy, University of Edinburgh, Peter Guthrie Tait Road,	Edinburgh EH9 3FD}

\ead{ivan.lobaskin@ed.ac.uk}
\ead{mevans@staffmail.ed.ac.uk}

\begin{abstract}
Totally asymmetric tracer particles in an environment of symmetric hard-core particles on a ring are studied. Stationary state properties, including the environment density profile and tracer velocity are derived explicitly for a single tracer. Systems with more than one tracer are shown to factorise into single-tracer subsystems, allowing the single tracer results to be extended to an arbitrary number of tracers. We demonstrate the existence of a cooperative effect, where many tracers move with a higher velocity than a single tracer in an environment of the same size and density. Analytic calculations are verified by simulations. Results are compared to established results in related systems. 
\end{abstract}
\section{Introduction}
Active matter systems comprise particles which consume energy in order
to perform work or generate motion. Their study lies at the heart of
intracellular biological physics where, for example, ATP conversion provides energy
to propel molecular motors and also nonequilibrium
statistical physics, where self-propulsion precludes equilibrium with
the environment \cite{R10,ND17}.

A  simple scenario is a single active particle in a bath
of otherwise equilibrium particles. One seeks to understand whether
close to equilibrium concepts such as fluctuation-dissipation theorems
will hold or whether the active particle can drive the whole bath far from
equilibrium.  One signature of a system being out of equilibrium is
that it exhibits physical currents, such as flow of particles, as well
as probability currents in phase space, reflecting lack of detailed
balance. Thus the question is whether the active particle can generate
a system-wide current.

A particularly simple model system for a bath of interacting particles is the Symmetric Simple  Exclusion Process (SEP). This comprises 
particles on a lattice moving stochastically to neighbouring sites but with hard-core repulsion interaction. The symmetry of the stochastic dynamics ensures that no current flows and detailed balance holds.
In contrast the Asymmetric Simple  Exclusion Process (ASEP)
has asymmetric stochastic dynamics which generate currents
and is a paradigmatic nonequilibrium system. The exclusion interaction in one dimension implies a no passing constraint which generates long range correlations. For a detailed review of the SEP/ASEP and its applications, see \cite{CMZ11}.

The problem of a single active ASEP particle---we will refer to it as the 
driven tracer particle (DTP)---in a background of 
SEP particles  has been studied extensively by Oshanin and co-workers \cite{BOMM92,BOMR96,BCLMO99,BCMO99,IBMOV13,BIOSV16,PBDO19}.
Initially  a totally asymmetric  DTP hopping with infinite rate on an infinite system of SEP particles was considered and it was shown
that the displacement of the DTP grows as $t^{1/2}$ with 
a prefactor given by a transcendental equation \cite{BOMM92}.
The results were extended to a partially asymmetric DTP  hopping with finite asymmetric rates \cite{BOMR96}
and simple forms for the prefactor were given in the weak and strong asymmetry regimes. 
The hydrodynamic limit and a law of large numbers for the tracer particle were proven rigorously in \cite{LOV98}.
In the limit of a high density background (i.e. density of SEP particles
approaching unity) all cumulants of the DTP displacement have been computed and shown to to scale as $t^{1/2}$ \cite{IBMOV13}.
Also a DTP has been studied in a related model defined on continuous space, the random average process, and its displacement also shown to have a $t^{1/2}$ scaling \cite{CKMM16}.
Further generalisations to  a background of particles
with fluctuating density due to desorption and absorption have been considered \cite{BCLMO99,BCMO99}.
Recently the problem of many partially asymmetric DTP's with different bias strengths  has been considered  and it has been shown how entrainment occurs,
for example if two DTP's  are biased in the same direction, they move faster than when they are alone and 
if they are biased in different directions, they eventually move in the direction of the stronger bias \cite{PBDO19}.

In this work, on the other hand, we will consider a finite periodic system
of size $L$ and compute the stationary state which is attained as $t\to \infty$ and the stationary properties such as the
DTP velocity.
To our knowledge the problem was first considered in a periodic  one-dimensional system in \cite{FGL85}.
It was shown that an Einstein relation holds which relates the velocity of 
a weakly asymetric DTP to the equilibrium diffusion constanst of a tagged SEP particle. Such an Einstein relation holds in all dimensions  \cite{HE93} for  a finite system when a perturbation generates a small current in an otherwise 
equilibrium system. It is known \cite{DM97} that the variance
of the displacement of a tagged SEP particle scales as $t/L$ therefore 
the velocity of a weakly asymmetric DTP should scale as $1/L$. A DTP in a SEP background has also been considered in \cite{CMP17,MMP19} as a simplification of a driven tracer in a narrow channel.
It was found that if the DTP is allowed to pass through the background particles at some rate, a nonequilibrium phase transition occurs between a sub-diffusive ``single file" phase with a vanishing DTP velocity and a diffusive ``ballistic" phase with a finite DTP velocity. We also mention that  fixed localised spatial defects that drive SEP particles have been studied in $d>2$ in \cite{SMM11,SMM14}.

Here we will consider  totally asymmetric DTP's  with hopping rate $p$.
By using a Matrix product formulation of the stationary state \cite{Evans96,BE07}
and a mapping to an inhomogeneous zero-range process \cite{Evans00,EH05} we show how the stationary state factorises
about the DTP's. This allows the density profiles of the background SEP particles about the DTP's to be computed.
We compute exactly the velocity of the tracer particles (for stationarity under exclusion all DTP's and background particles  necessarily have equal velocity) and show that it has the expected $1/L$ scaling with a prefactor which depends on the background density. We demonstrate how for several tracers
the entrainment effect enhances the velocity.

The paper is organised as follows. In Section \ref{defn} we define the model in the case of a single DTP. In Section \ref{zrp}  we use the matrix product formalism and a mapping to a zero-range process to obtain expressions for the partition function of the steady state of the single DTP system. In Section \ref{pf} we calculate the generating function of this partition function, which allows us to evaluate it explicitly. In Section \ref{dp} we use these results to calculate the density profile of the bath and stationary velocity of the DTP. In Section \ref{many} we examine systems with many DTP's and show that their partition functions can be factorised into single DTP subsystems. This allows us to extend the results from Section \ref{dp} to many DTP systems and we conclude by examining the effects of the presence of multiple DTP's on the stationary velocity of the system.

\subsection{Definition of single DTP model}\label{defn}
We consider a one-dimensional periodic lattice with $L$ sites. On it we place $M$ particles which interact by simple exclusion. $M-1$ of these are ``bath" particles, which hop to the left and to the right at rate $1/2$, and the last particle is the DTP, which hops only to the right at rate $p$. Then we have $N=L-M$ empty sites and an average density of $\rho=M/L$.

We are interested in how the presence of the DTP will affect the distribution of particles in the system. From translational symmetry, it follows that in the stationary state, the density of particles will be uniform. A more illuminating perspective is to move to the reference frame of the DTP. There the system settles into a nonequilibrium stationary state with a non-uniform density profile.
In this reference frame, the DTP will always be at site $0$ and the rest of the system will consist of $M-1$ symmetrically hopping particles and $N$ empty sites. The hops of the DTP (to the right) instead become simultaneous hops of the bath particles to the left.

\begin{figure}[h]
	\centering 
	\includegraphics[scale=0.25]{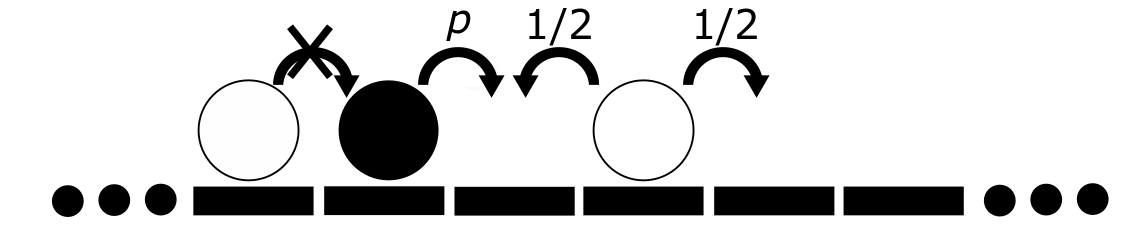}
	\caption{Simple exclusion process with symmetrically hopping bath particles (unfilled) and asymmetrically hopping tracer (filled)}
\end{figure}

\section{Zero-range process}\label{zrp}
In order to find the steady state it is helpful to use the well known mapping in one dimension from exclusion process  onto a zero-range process \cite{EH05}. The mapping  in the case of a single driven particle has previously been used in \cite{CMP17} where  the velocity
and density profile were calculated in the weakly asymmetric limit.
 
The mapping is as follows. We associate to our original lattice (SEP) with $M$ particles and $L$ sites  a different lattice (ZRP) with $M$ sites and $N$ particles. We assign each site in the ZRP lattice to one particle in the SEP lattice. Then we place particles on each ZRP site equal to the number of empty sites in front of the corresponding particle in the SEP lattice. Thus if there are $n_1$ empty sites in front of the DTP, then there are $n_1$ particles on the first ZRP site; if there are $n_2$ empty sites in front of the first bath particle in the SEP, there are $n_2$ particles on the second ZRP site etc.

The name zero-range process comes from the condition that the rate at which particles hop out of a site depends only on the departure site. In the present case, this is trivially true as the rate of hopping in the ZRP picture 
depends only on the location of the departure site:
$p+1/2$ for the first site, $1/2$ for the last site and $1$ for all other sites. As these rates vary by location, we refer to it as an inhomogeneous ZRP.

\begin{figure}[h]
	\centering
	\includegraphics[scale=0.2]{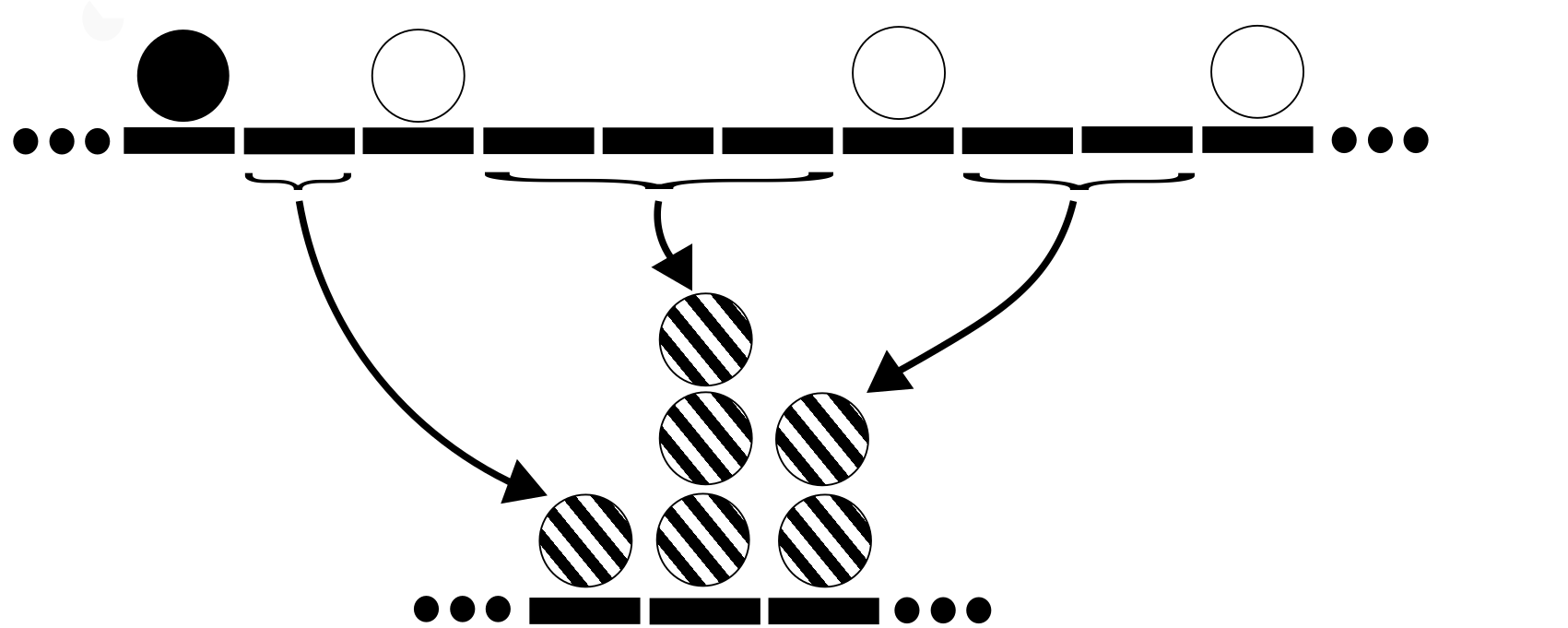}
	\caption{Mapping of simple exclusion process to zero-range process. The ZRP particles (striped) correspond to the number of empty sites in front of tracer (filled) and bath particles (unfilled) in SEP}
\end{figure}

\subsection{Factorisation of steady state}
The advantage of working in the ZRP lattice is that the steady state of the system can be written in a factorised form \cite{Evans00}. Let $n_i$ be the occupation of the $i$-th ZRP site. Then we can write the probability of the state $\{ n_1,n_2,\dots ,n_M\}$ as:
\begin{equation}
	P(\{ n_1,n_2,\dots ,n_M\} ) \propto f_1(n_1)f_2(n_2)\dots f_M(n_M)
	\label{Pf}
\end{equation}
for some functions $f_1,f_2\dots , f_M$. We can also write down a (canonical) partition function:
\begin{equation}
	Z_{M,N} = \sum\limits_{\{ n\} } \prod\limits_{i=1}^{M}f_i(n_i) \label{factor}
\end{equation}
where $\{ n\}$ denotes that the sum is over all configurations $\{ n_1,\dots , n_M\}$ of $N$ identical particles on $M$ sites.

To find an explicit form for the functions $f_i$, we note the stationarity condition  which comes from equating the rate of leaving a configuration
with the rate of entering it,
\begin{eqnarray}
	\lefteqn{\left[ (p+\frac{1}{2})\theta(n_1) + \sum_{i=2}^{M-1}\theta(n_i) + \frac{1}{2} \theta(n_M)\right]
P(\{ n_1,n_2,\dots,n_M\}) } \nonumber \\
 	&=& \frac{1}{2}\theta(n_1) P(\{ n_1-1,n_2+1,\dots,n_M\} )\nonumber  \\
	&& + \sum_{i=2}^{M-1}\frac{\theta(n_i)}{2}\left[
	P(\{ \dots n_{i-1}+1,n_i-1, \dots \} ) 
	+ P(\{ \dots n_i-1,n_{i+1}+1, \dots \} )\right] \nonumber \\
	&& + \theta (n_M) pP(\{ n_1+1, \dots ,n_{M-1},n_M-1\} ) \label{p3}
\end{eqnarray}
where
\begin{equation}
\theta(n) =
\cases{
	1 \quad \mbox{if} \quad n>0\\
    0 \quad \mbox{if} \quad n=0\;.
}
\end{equation}
Now equating term prefixed by each $\theta(n_i)$ and  defining 
\begin{equation}
g_i\equiv \frac{f_i(n_i+1)}{f_i(n_i)}\;, \label{gdef1}
\end{equation}
yields the conditions
\numparts
\begin{eqnarray}
	(p+1/2)g_1 & = & \frac{1}{2}g_2 \\
	g_i & = & \frac{1}{2} ( g_{i-1} + g_{i+1} )\\
	\frac{1}{2}g_{M} & = & \frac{1}{2}g_{M-1} + pg_1\;.
\end{eqnarray}
\endnumparts
We assume $g_i$ to be independent of $n_i$, which implies
\begin{equation}
	f_i(n_i)=g_i^{n_i}\;. \label{fg}
\end{equation}
This  allows us to solve the system of equations and obtain the result
\begin{equation}
	g_i = g_1(1+2p(i-1))\;. \label{gdef2}
\end{equation}
Noting that $g_1$ always appears exactly $N$ times in the partition function, we can set it to $1$ without loss of generality. Then the partition function becomes
\begin{equation}
	Z_{M,N} = \sum\limits_{\{ n\} } \prod\limits_{i=1}^{M} (1+2p(i-1))^{n_i}\;. \label{factorp}
\end{equation}

\subsection{Matrix product state}
The steady state can also be represented as a matrix product \cite{BE07}
as we now discuss.  This approach turns out to be more convenient for deriving the density profile in the steady state.

It was shown in \cite{Evans96} that the stationary probabilities of  one dimensional periodic exclusion process
in which each particle $\mu$ has its own hopping rates $p_\mu$ and $q_\mu$ to the right and left respectively can be written in Matrix product form.
In the process we consider here, which involves a single driven tracer particle,
we write the stationary weights as the trace of a product of matrices corresponding to the occupancy of each site:
the DTP corresponds to a matrix $B$, the bath particles to matrix  $D$ and the empty sites to a matrix $E$. 

Then the probability of the configuration $\{ n_1, \dots ,n_M\}$ corresponds to the matrix product
\begin{equation}
	P(\{ n_1, \dots ,n_M\} ) \propto \Tr [ BE^{n_1}D\dots DE^{n_M} ] 
\end{equation}
where the trace implies that there is translational invariance with respect to the position of the DTP. For this matrix product to satisfy the stationary 
master equation we require that these matrices satisfy the conditions:
\numparts
\begin{eqnarray}
	BE &=& B \label{be}\\
	DE &=& ED + 2pD \label{de} \;.
\end{eqnarray}
A general proof is given in \cite{BE07} which we do not repeat here.

These conditions  in turn generate reduction rules which allow the matrix product to be reduced. 
As a check that we obtain the same result as through the mapping to the zero-range process, we note first a consequence of (\ref{de})
\begin{equation}
	D(E+ a)  =(E+ a + 2p)D \label{de2} \;.
\end{equation}
\endnumparts
where $a$ is a scalar.
Then it is apparent using (\ref{be},\ref{de2}) that
\begin{eqnarray}
\Tr [ BE^{n_1}D E^{n_2}\dots DE^{n_M} ] &=&
\Tr [ BE^{n_1}(E+2p)^{n_2} \ldots (E+2p(M-1))^{n_M}D^M ]\nonumber  \\
&=&
\prod\limits_{i=1}^{M} (1+2p(i-1))^{n_i}\,  \Tr [ BD^M ]
\end{eqnarray}
and $\Tr [ BD^M ]$ is just a constant factor that can be set to unity without loss of generality.

The partition function can then be written as a matrix product as:
\begin{equation}
	Z_{M,N} = \{ z^N\} \Tr [BC^{L-1}] \label{zmat}
\end{equation}
where we have introduced the matrix $C\equiv D+zE$, $z$ is an auxiliary variable (the fugacity) and the notation $\{ z^N \}$ signifies that we only take the coefficient of $z^N$ of the expression that follows. 

\section{Calculation of the partition function} \label{pf}
The partition function in the form given in (\ref{factorp}) is impractical for calculations. To obtain a more useful expression, we will first show how to express the partition function in integral form using the generating function method and then we will proceed to calculate the saddle point of the integral, which will be used in section \ref{dp} to derive the density profile.
\subsection{Generating function method}
First we define the generating function (or grand canonical partition function). Let $z$ be an auxiliary variable, then the generating function is defined as:
\begin{equation}
	{\cal Z}_M(z) \equiv \sum\limits _{n=0}^\infty z^n Z_{M,n}\;.
	\label{gf}
\end{equation}
In the case $M=1$, $Z_{M,N}=1$ for any $N$ since there is only one possible state. Hence
\begin{equation}
	{\cal Z}_1(z) = \sum\limits _{n=0}^\infty z^n =\frac{1}{1-z}\;. \label{z1}
\end{equation}
To calculate the generating function for all $M$, we proceed by induction. Consider the system in the ZRP picture. 
The (canonical) partition function of the system with $M$ sites can be obtained by multiplying the partition function of a system with $M-1$ sites by the weight of the $M$-th site, $g_M$, as given by equation (\ref{gdef2}), and summing over all possible number of particles ($n=0,1,\dots ,N$) in the $M$-th site,
\begin{equation}
	Z_{M,N} = \sum\limits _{n=0}^N Z_{M-1,N-n}g_M ^n\;.
\end{equation}
By substituting this into (\ref{gf}) and manipulating the sums, we obtain a recursion relation for the generating function:
\begin{equation}
	{\cal Z}_M (z)=
	 {\cal Z}_{M-1}(z)\sum _{n=0} ^\infty (zg_M)^n \;,
\end{equation}
which combined with (\ref{z1}) yields the result

\begin{equation}
	{\cal Z}_{M}(z) = \prod\limits _{j=1}^M \frac{1}{1-g_jz}\;.
\end{equation}
Then the canonical partition function can be written using the Cauchy integration formula as:
\begin{equation}
	Z_{M,N} = \frac{1}{2\pi i}\oint dz \ z^{-(N+1)}
	\prod\limits _{j=1}^M \frac{1}{1-g_jz}
\end{equation}
where the contour encloses the origin.
\subsection{Saddle point calculation}
The exact form of this integral is not very tractable but in the thermodynamic limit, $M,N \rightarrow \infty$, it can be approximated very well using the saddle point method. First we rewrite the partition function in a form that is more convenient for a saddle point calculation
\begin{equation}
	Z_{M,N} = \frac{1}{2\pi i}\oint \frac{dz}{z} \ \exp
	\left( -N\log z - \sum\limits _{j=1}^M \log (1-g_j z) \right) \;. \label{zint}
\end{equation}
To turn this into a form amenable to standard saddle point approximation 
we change variable to  $\zeta \equiv 2p M z$.
Noting that $g_j \sim j$, the sum will be dominated by the terms with $j\approx M$. Then we can make the approximation $g_j=1+2p(j-1)\approx 2pj$.  Finally, we define $y \equiv j/M$ and replace the sum with an integral, obtaining
\begin{equation}
	Z_{M,N} = \frac{(2pM)^N}{2\pi i}\oint \frac{d\zeta}{\zeta} \exp
	\left( -N\log \zeta - M\int\limits _{0}^1 dy \log (1-y \zeta) \right)\;.
\end{equation}
Evaluating the $y$ integral, we end up with
\begin{equation}
	Z_{M,N} = \frac{(2pM)^N}{2\pi i}\oint \frac{d\zeta}{\zeta} \exp 
	\left\{ -M\left[ \frac{1-\rho}{\rho}\log \zeta +
	\left( 1-\frac{1}{\zeta} \right) \log (1-\zeta ) +1 \right] \right\} \;.
	\label{ZLNint}
\end{equation}
where we have used $\displaystyle \frac{N}{M}=\frac{1-\rho}{\rho}$. In this form, it is obvious that the integral will be dominated by the saddle point, located at the turning point of the function inside the exponential. Setting its derivative to zero, we find that the saddle point is located at $\zeta _0$, which is implicitly defined as the positive solution of the transcendental equation
\begin{equation}
	\zeta _0 = 1-{\rm e}^{-\zeta _0/\rho} \;.\label{xi}
\end{equation}
We specify ``positive" as the equation always has the trivial solution of 0, but this is irrelevant for the saddle point. $\zeta _0$ can be expressed in terms of special functions or simply found numerically.
Then evaluating the integral gives us the expression
\begin{equation}
	Z_{M,N}	\simeq 
	A(\rho)
	\exp\left\{ -N\log \left( \frac{\zeta_{0}}{2pM}\right) 
	-\frac{M}{\rho }(1-\zeta_{0})+M\right\} \label{ZLN}
\end{equation}
where
\begin{equation}
	A(\rho) = \left[ 2\pi\left( \frac{1}{1-\zeta _0 }-\frac{1}{\rho}\right) \right] ^{-1/2}
\end{equation}
and (\ref{ZLN}) has corrections that are diminished  by a factor $O(1/M)$.

\section{Density profile}\label{dp}
\begin{figure}
	\centering
	\includegraphics[scale=0.8]{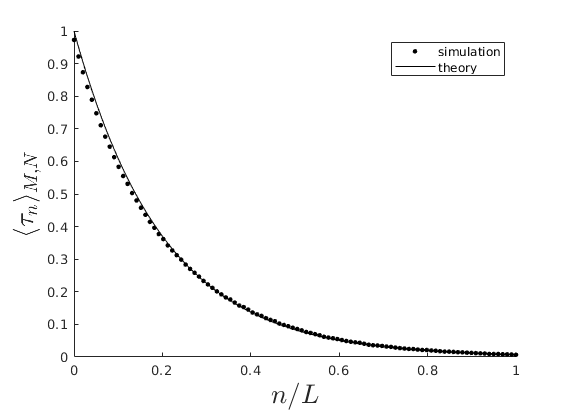}
	\caption{Density profile from theory and simulations for a system with $p=1$, $M=20$, $L=100$ ($\rho = 0.2$). The slight discrepancy visible near $n=0$ is due to $O(1/M)$ corrections.}
	\label{density_profile}
\end{figure}

To calculate the density profile, we use the matrix product formalism. Using the expression for the partition function (\ref{zmat}), we can get the average occupancy of $n$-th site, $\langle \tau _n\rangle _{M,N}$, by replacing the $n$-th $C$ with a $D$ (which, as a reminder, stands for a filled site). 
\begin{equation}
	\langle \tau _n \rangle _{M,N} = 
	\frac{1}{Z_{M,N}}\{ z^N \} \Tr [BC^{n-1}DC^{L-1-n}]	\;.
\end{equation}
From the relations (\ref{be},\ref{de}) it follows that $CD=D(C-2pz)$. Then the single $D$ can be commuted with all $C$s to its left to give
\begin{equation}
	\langle \tau _n \rangle _{M,N} = 
	\frac{1}{Z_{M,N}}\{ z^N \} \Tr [BD(C-2pz)^{n-1}C^{L-1-n}]\;.
\end{equation}
Now, performing the binomial expansion, we will obtain a sum of terms of the form $(-2p)^k\{ z^{N-k}\}\Tr [BDC^{L-k-2}] $, which are proportional to the average occupancy of the first site in a lattice with $M$ particles and $N-k$ empty sites. Thus we obtain
\begin{equation}
	\langle \tau _n \rangle _{M,N} = 
	\frac{1}{Z_{M,N}} \sum\limits _{k=0}^{n-1} {n-1 \choose k}
	(-2p)^k Z_{M,N-k}\langle \tau _1 \rangle _{M,N-k}\;.
\end{equation}
We show in section \ref{site1} that $\langle \tau _1 \rangle _{M,N-k}=1-O(1/M)$. To be exact, $\langle \tau _1 \rangle _{M,N-k}$ should be set to $0$ for $k>N$. However, it can be shown that the final result is dominated by  the term with $k\approx \xi _0/\rho M\ll N$, so the error introduced by ignoring this is negligible. We can now use the integral form of the partition function (\ref{ZLNint}) to rewrite this expression as
\begin{equation}
	\langle \tau _n \rangle _{M,N} = 
	\frac{1}{Z_{M,N}}\frac{1}{2\pi i}\oint \frac{dz}{z^{N+1}}(1-2pz)^{n-1} {\cal Z}_M(z)\;.
\end{equation}
A simple analysis shows that the extra factor $(1-2pz)^{n-1}$ does not change the location of the saddle point, for all values of $n$. Then the integral simply becomes $(1-2pz_0)^{n-1}Z_{M,N}$, where $z_0$ is the saddle point. Hence we get
\begin{equation}
	\langle \tau _n \rangle _{M,N} = 
	\exp \left( -\frac{\zeta _0}{\rho} \frac{n-1}{L} \right)\;.
	\label{profile}
\end{equation}
Thus we obtain an exponential profile with decay length $\frac{\rho }{\zeta _0}L$, where $L$ is the size of the whole system. This is found to be in very good agreement with simulations (see Fig.~\ref{density_profile}).
\subsection{Occupation of the first site and DTP velocity}\label{site1}
We now explicitly calculate $\langle \tau _1\rangle _{M,N}$. We begin by noting that the probability that the first site is empty, $1-\langle \tau _1\rangle _{M,N}$, can be simplified using the relation (\ref{be}) as follows:
\begin{equation}
	1-\langle \tau _1\rangle _{M,N} = 
	\frac{\{ z^{N-1}\} \Tr [BEC^{L-2}]}{Z_{M,N}}
	= \frac{Z_{M,N-1}}{Z_{M,N}}\;.
\end{equation}
From (\ref{ZLN}, \ref{zint}) we see that this ratio of partition functions will be given by the saddle point value of $z$, which is the fugacity in the grand canonical ensemble. Thus, to leading order in $1/M$, we get the very simple expression
\begin{equation}
	1-\langle \tau _1\rangle _{M,N} = \frac{\zeta _0}{2pM}+O(M^{-2})
\end{equation}
which validates the approximation $\langle \tau _1\rangle _{M,N}=1+O(1/M)$ above. This also allows us to find the stationary velocity of the tracer $\langle v _T \rangle _{M,N}$ (and therefore the whole system):
\begin{equation}
	\langle v_T \rangle _{M,N}
	= p(1-\langle \tau _1\rangle _{M,N}) 
	= \frac{\zeta _0}{2M}=\frac{\zeta _0}{2\rho}\frac{1}{L}\;.
	\label{v}
\end{equation}
Thus we get the expected scaling $\langle v\rangle \sim 1/L$. We remark that this result does not depend on $p$. The stationary velocity is controlled only by how quickly the symmetric particles can diffuse away from the DTP. 
\section{Many DTP's}\label{many}
We now turn to systems with more than one DTP. It turns out that in this case the partition function can be factorised into partition functions of single DTP subsystems. We cover the case with two DTP's explicitly and then extend this argument to an arbitrary number.
\subsection{Two DTP's} \label{2tasep}
Let the number of empty sites in the system be $N$ and let there be $M_1+M_2$ particles in the system. The particles with label $1$ and $M_1 +1$ are DTP's, both hopping with rate $p$ to the right and the rest are bath particles, as before. We can again perform the mapping to the zero-range process and obtain an equation analogous to (\ref{p3}). We find that the solutions for $i\leq M_1$ are the same as in the single DTP case. Then we get a ``reset", with $g_{M_1+1}=1$, and the $g_i$ start increasing by $2p$ again:
\numparts
\begin{eqnarray}
		g_i &=& 1+2p(i-1), \ 1\leq i\leq M_1 \\
		g_{M_1+i} &=& 1+2p(i-1), \ 1\leq i\leq M_2\;.
\end{eqnarray}
\endnumparts
This is readily seen in the matrix product formulation where, for example,
\begin{eqnarray}
\lefteqn{\Tr [ BE^{n_1}D\dots DE^{n_{M_1}}BE^{n_{M_1}+1}D\dots DE^{n_{M_1+M_2}}B  ]}\nonumber\\
 &=&
\Tr[ BE^{n_1} \ldots (E+2p(M_1-1))^{n_{M_1}}D^{M_1}
BE^{n_1} \ldots (E+2p(M_2-1))^{n_{M_2}}D^{M_2}\nonumber ]\\
&=&
\prod\limits_{i=1}^{M_1} (1+2p(i-1))^{n_i}\,
\prod\limits_{j=1}^{M_2} (1+2p(j-1))^{n_{M_1+j}}\,
  \Tr [ BD^{M_{1}}BD^{M_{2}} ]\;.
\end{eqnarray}
Writing down the partition function similarly to (\ref{factorp}), we notice that due to the factorised form of the steady state, it is simply a sum over products of two partition functions of systems with one DTP:
\begin{equation}
	Z_{M_1,M_2,N} = \sum\limits _{n=0}^N Z_{M_1,n}Z_{M_2,N-n}\;. \label{Z2L}
\end{equation}
The sum runs over all possible way to divide $N$ empty sites between the two subsystems: term $n$ corresponds to the configuration with $n$ empty sites in subsystem 1 and $N-n$ empty sites in subsystem 2. In the large $M_1,M_2$ limit, we can substitute (\ref{ZLN}) for the single DTP partition functions to obtain
\begin{equation}
	Z_{M_1,M_2,N} = 
	\sum\limits _{n=0}^N
	A(\rho _1)A(\rho _2)\exp (-b(M_1,M_2,N,n))
	\label{eq_zmm}
\end{equation}
where $\rho _1=M_1/(M_1+n)$, $\rho _2=M_2/(M_2+N-n)$ and $b(M_1,M_2,N,n)$ is the function obtained by adding the arguments of the exponential in (\ref{ZLN}) for two single DTP systems with parameters $M_1,n$ and $M_2,N-n$. Noting that $b\sim M_1,M_2$, we expect this sum to be sharply peaked around a single term, which corresponds to the most likely division of empty sites in the steady state. We check the validity of this approximation at the end of this section, when we calculate the subleading term. Let the location of the dominant term be $n_0$, which we use to define the stationary densities of the two subsystems $\rho ^* _1=M_1/(M_1+n_0)$ and $\rho ^* _2=M_2/(M_2+N-n_0)$. Setting $\partial _n b=0$, we find that $n_0$ is given implicitly by the equation

\begin{equation}
	\zeta _1/M_1 = \zeta _2/M_2 \;.\label{z1z2}
\end{equation} 
where $\zeta _{1,2}$ correspond to the solutions of (\ref{xi}) with densities $\rho ^*_{1,2}$. Using (\ref{v}) we see that this condition simply states that the densities are such that the two subsystems share the same fugacity, which implies equal velocities for the two DTP's. We now have a system of four equations for the unknowns $\rho ^* _1,\zeta _1, \rho ^* _2,\zeta _2$, namely: equation (\ref{z1z2}); equation (\ref{xi}), which is satisfied both by $(\rho ^*_1,\zeta _1)$ and $(\rho ^*_2,\zeta _2)$; and finally the condition that the number of empty sites in the two subsystems always sums to $N$, which we can write as:
\begin{equation}
	\frac{M_1}{\rho ^*_1}+\frac{M_2}{\rho ^*_2}=L\;.
\end{equation}
We now denote the common stationary velocity as $\langle v_T \rangle _{M_1,M_2,N}= \zeta _1/2M_1=\zeta _2/2M_2$ and define
\begin{equation}
	\Xi _0 \equiv 2(M_1+M_2)\langle v_T \rangle _{M_1,M_2,N}\;. \label{Xidef}
\end{equation}
We can reduce the system of equations to an analogue of (\ref{xi}) for $\Xi _0$:
\begin{equation}
	(1-m_1 \Xi _0)(1-m_2 \Xi _0)={\rm e}^{-\Xi _0/\rho} \label{Xi}
\end{equation}
where $m_{1,2}=M_{1,2}/(M_1+M_2)$ are fractions of particles in the two subsystems, and $\rho=(M_1+M_2)/L$ is the average density of the whole system. From (\ref{Xidef}) we have the relation $\Xi _0 = \zeta _{1,2}/m_{1,2}$. Since $\zeta _{1,2}$ must satisfy (\ref{xi}), we have $\zeta _{1,2}<1$, so $\Xi _0 < 1/m_{1,2}$. It can be shown that (\ref{Xi}) has a unique solution for $0<\Xi _0<\min (1/m_1,1/m_2)$.

We now have a recipe for calculating the stationary velocity and the densities of the subsystems, $\rho ^* _1,\rho ^*_2$. Given the parameters $M_1,M_2,N$, we can solve (\ref{Xi}) numerically for $\Xi _0 $. This immediately gives us the stationary velocity through (\ref{Xidef}). The stationary velocity of two DTP's, scaled to that of one DTP, is plotted in Fig.~\ref{speed_fig} as a function of $m_1$ at various densities. This illustrates that the velocity of more than one DTP is always higher than that of a single DTP. We will discuss this cooperative effect further in Section~\ref{fast_part}. Knowing $\langle v_T \rangle _{M_1,M_2,N}$, we can get $\zeta _{1,2}$, which we can then substitute into (\ref{xi}) to find the densities $\rho _ {1,2}$. Thus we can approximate the combined partition function as
\begin{equation}
	Z_{M_1,M_2,N} \approx Z_{M_1,M_1(1-\rho ^*_1)/\rho ^*_1}Z_{M_2,M_2(1-\rho ^*_2)/\rho ^*_2}
\end{equation}
where $\rho ^*_{1,2}$ are calculated with the prescribed recipe. Since the partition functions of the subsystems are those of one DTP systems, the density profiles in front of the DTP's are exponential. Then (\ref{z1z2}) means that the two profiles have the same characteristic lengths (but the total lengths of the two subsystems are generally different).

We now check the validity of the leading order approximation by calculating the subleading term in $n-n_0$ (where $n$ is the number of empty sites in the first subsystem). Close to $n_0$, we have $b(n)\approx b(n_0)+\frac{1}{2}(n-n_0)^2b''(n_0)$, where $b$ is as in (\ref{eq_zmm}) and $'$ denotes a derivative with respect to $n$. We find
\begin{equation}
	b''(n_0) =
	\sum\limits _{i=1,2}
	\frac{1}{M_i}\frac{\rho ^*_i (1-\zeta _i)}{\rho ^*_i+\zeta _i-1}
	\;. \label{nvar}
\end{equation}
Note that the probability of observing a configuration with a particular value $n$ is
\begin{equation}
	P(n) = \frac{Z_{M_1,n}Z_{M_2,N-n}}{Z_{M_1,M_2,N}}\;.
\end{equation}
We can approximate this as a Gaussian in $n$ with variance
\begin{equation}
	\sigma _n ^2 = 
	[b''(n_0)]^{-1}
	\sim M_i \;. 
\end{equation}
So the relative fluctuations of $n$ scale as $M_i^{-1/2}$ and vanish in the thermodynamic limit. The variance of $n$ has also been estimated through simulations, by tracking the relative positions of the two DTP's, and was found to be in excellent agreement with (\ref{nvar}).
\subsection{$k\geq 2$ DTP's}
The calculations from section \ref{2tasep} are straightforward to extend to a case of $k\geq 2$ DTP's. We now divide the system into $k$ subsystems, with $M_1,M_2,\dots ,M_k$ particles (the first particle in each being a DTP) and $N$ empty sites. We consider the limit where all $M_i$ are large but $k$ is fixed. As before, the partition function can be written as a sum of products of single DTP partition functions. Then if there are $n_1,n_2,\dots ,n_k$ empty sites in the subsystems (with $n_1+n_2+\dots +n_k = N$), finding the largest term with respect to $n_1,n_2,\dots $, we get the equal fugacity condition
\begin{equation}
	\zeta _i /M_i = \zeta _j/M_j,\ \forall i,j
\end{equation}
which can be reduced to
\begin{equation}
	\prod\limits _{i=1}^k (1-m_i \Xi _0) = {\rm e}^{-\Xi _0/\rho}
	\label{Xik}
\end{equation}
where $m_i,\Xi _0$ are the obvious generalisations of the definitions in \ref{2tasep}. Then to leading order we can factorise the partition function into single DTP partition functions (with subsystem densities $\rho ^*_i$ calculated using the same procedure as before):
\begin{equation}
	Z_{M_1,M_2,\dots ,M_k,N} \approx 
	\prod\limits _{i=1}^k Z_{M_i,M_i(1-\rho ^*_i)/\rho^*_i}
	\;.
\end{equation}
We get a series of subsystems with equal stationary velocities and exponential density profiles with equal decay rates.
\subsection{Effect of many DTP's on stationary velocity} \label{fast_part}
\begin{figure}[h]
	\centering
	\includegraphics[scale=0.8]{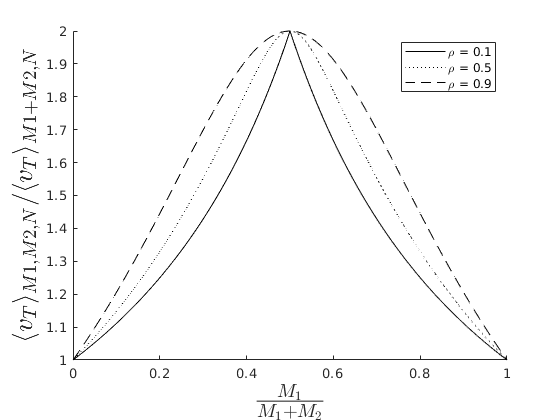}
	\caption{Ratio of stationary velocity of two DTP's to that of one DTP at various densities. There is a clear peak when $M_1=M_2=M/2$ and the minimum of $1$ is achieved when $M_1$ or $M_2$ is $0$.}
	\label{speed_fig}
\end{figure}

We now examine how the stationary velocity is affected by the presence of many DTP's and specifically how it compares to the single DTP case. Although it is difficult to analyse (\ref{Xik}) directly, we can look at extremal cases. First we investigate the following question: given a fixed system size $N,M$, which partition into $k$ subsystems $\{ M_1,M_2,\dots ,M_k\}$ (where $M_i$ is the number of particles in subsystem $i$) gives the highest stationary velocity? Noting that $\Xi _0$ is proportional to the stationary velocity, we maximise $\Xi _0$ with respect to all $m_i$ subject to the constraint $\sum _{i=1}^k m_i =1$. This gives us the following condition for the maximum:
\begin{equation}
	m_1^*=m_2^*=\dots =m_k^*=1/k\;.
\end{equation}
So the fastest partition is when the DTP's divide the system into equal parts. Putting these values into (\ref{Xik}), we get
\begin{equation}
	\Xi _0 ^* (\rho)= k\zeta _0 (\rho )
\end{equation}
where $\zeta _0(\rho )$ is the solution of (\ref{xi}). This means that in the optimal scenario, the $k$ DTP system moves at $k$ times the velocity of a one DTP system of the same size and density. In the opposite extreme, if (without loss of generality) $m_1=1$ and $m_i=0$ for $i\neq 1$, we find that equation (\ref{Xik}) reduces exactly to (\ref{xi}) and the stationary velocity of the $k$ DTP system is the same as that of a one DTP system of the same size and density. Thus the effectiveness of the cooperation depends on how the system is partitioned by the DTP's. The ratio of the velocity of a $k=2$ DTP system to that of a single DTP system is plotted in Fig.~\ref{speed_fig}. The predicted extremal values can be seen and also that a system of many DTP's always moves at least as fast as a single DTP system.
\section{Conclusions}
We have used a mapping to a zero-range process and the matrix product formalism to perform exact calculations for a hard-core lattice gas with driven tracer particles (DTP) in a periodic one-dimensional system. We found that the density profiles in front of the DTP's are exponential with characteristic lengths typically on the scale of the system size, which shows that a single driven particle can in fact create a system wide current in a finite system. We also found that the stationary velocity of the system scales as $1/L$, where $L$ is the size of the system.

For the case of many DTP's, we showed that the steady state can be factorised into single DTP subsystems. In each subsystem, the density profile decays exponentially. We also found that many DTP's can cooperate and achieve a velocity greater than that of a single DTP, though the extent of this effect depends on how the DTP's are placed in  the system. These results for the one and many DTP cases were found to be in very good agreement with simulations.

Our system-sized exponential density profile is similar to the result obtained in the weakly asymmetric tracer case \cite{CMP17}. In related infinite models, in particular models with desoprtion \cite{BCLMO99}, models in higher dimensions \cite{BCCMO01} and quasi one dimensional ``narrow channels" \cite{BIOSV16}, exponential density profiles were also observed in front of the tracer but with a finite decay length. Those models also exhibit a depletion zone behind the tracer, which was not present in the periodic systems studied in this work.

The effect of cooperation between many DTP's was also observed in infinite systems \cite{PBDO19}, where it was found that in the high density limit, many DTP's move as one and the effective force on the centre of mass is simply the sum of the forces on the individuals.

One could generalise our approach  to the case of partially asymmetric tracers in which case the matrix product approach still holds but with
each DTP (labelled $\mu$) represented by its own matrix $B_\mu$ \cite{Evans96}. In the general case one still has a factorised stationary state  (\ref{Pf}) with single site weights given by (\ref{fg}) but now the $g_i$ become interdependent. It would be of interest to investigate further how these interdependencies affect the stationary velocity.

Finally, one way of connecting the large time displacement in the infinite system and the finite $L$ systems we have studied here is through the scaling ansatz
\begin{equation}
\langle X(t)\rangle = t^{1/2} h(t/L^z)
\end{equation}
where here $z$ is the dynamic exponent, $X$ is the displacement of the DTP and $h(y)$ is a scaling function
which approaches a constant as $y \to 0$ to yield the infinite system scaling. In the opposite limit of $y\to \infty$ ($t \to \infty$ on a large but finite system)
we expect $h(y) \sim y^{1/2}$ so that we obtain a stationary velocity. Then we find  that the velocity $v_T \sim L^{-z/2}$ implying dynamic exponent $z=2$, which is the usual SEP behaviour. It would be of interest to calculate dynamical properties exactly.

\section*{Note added}
After completion of this work, we became aware of a preprint by Ayyer \cite{A20} which also derives the density profile induced by a driven tracer in a periodic hard-core lattice gas. The results regarding the density profile agree with ours but instead of using the mapping to zero-range process and matrix product formalism, he exploits combinatorial identities involving Stirling numbers. Ayyer also generalises this model to a partially asymmetric tracer but does not consider the case of more than one tracer.

\section*{Acknowledgements}
Ivan Lobaskin acknowledges studentship funding from EPSRC under Grant No. EP/R513209/1.

\section*{References}
\bibliographystyle{iopart-num}
\bibliography{references}

\end{document}